\documentclass[12pt]{iopart}

\usepackage{graphicx}

\newcommand{ \be }{\begin{equation}}       
\newcommand{ \ee }{\end{equation}}       
\newcommand{ \bea }{\begin{eqnarray}}       
\newcommand{ \eea }{\end{eqnarray}}

\newcommand{ \eps }{\varepsilon}       
       
\newcommand{ \mean }[1]{\left\langle #1 \right\rangle}   
   
\def\prl#1#2#3{ #3 {\it Phys. Rev. Lett.} {\bf #1} #2}

\begin{document}

\title[$v_2/\eps$ scaling]{Energy and system size dependence of
 charged  particle  elliptic flow and  $v_2/\eps$  scaling
}

\author{Sergei A. Voloshin
for the STAR Collaboration\footnote{For the full list of STAR authors 
and acknowledgments, see appendix `Collaborations' of this volume} 
}

\address{Wayne State University, Detroit, Michigan} 
\ead{voloshin@wayne.edu}


\begin{abstract}
  We report measurements of charged particle elliptic flow 
  at mid-rapidity in Au+Au and Cu+Cu collisions at
  $\sqrt{s_{_{NN}}}=62$~and 200 GeV. 
  Using correlations between main STAR TPC and Forward TPCs ensures
  minimal bias due to non-flow effects.
  We further investigate the effect of flow fluctuations on $v_2/\eps$
  scaling studying initial geometry eccentricity fluctuations 
  in Monte-Carlo Glauber model, consistent with STAR direct measurements of
  elliptic flow fluctuations~\cite{sorensenQM2006}.
  It is found that accounting for the effect of flow fluctuations 
  improves $v_2/\eps$ scaling. 
\end{abstract}



The large elliptic flow observed at RHIC~\cite{star-flow1}, along with its
mass dependence at low transverse momenta~\cite{star-flow2,v2pt-mass}
is indicative of early thermalization of the
system created in high energy nuclear collisions and its evolution
consistent with ideal hydrodynamics. 
The observed~\cite{star-flow-cqs} constituent quark
scaling~\cite{voloshin-qm2002,voloshin-molnar} 
of elliptic flow suggests that the system spends significant
time in the deconfined state.
Further insight into the physics of the elliptic flow and the
processes governing the evolution of the system can be achieved by
the study of the elliptic flow dependence on the system size and
collision energy. 
In~\cite{voloshin-poskanzer-PLB} the authors suggested that the
elliptic flow follows a simple scaling in the initial system
eccentricity and the particle density in the transverse plane, 
$v_2/\eps \propto 1/S~dN_{ch}/dy$
(where $S$ is the area of the overlap region of two nuclei). 
Indeed the available data are consistent with such a
scaling~\cite{voloshin-qm2002,na49-flow-PRC}. 
Unfortunately, the systematic uncertainties in these results 
are too large to conclude on how well the scaling holds. 
The main difficulties are the evaluation/elimination  of the
so-called non-flow correlations (azimuthal correlations not related to
the reaction plane orientation), effects of flow fluctuations,
and uncertainties in the calculation of the initial
system eccentricity.  

A technique to suppress non-flow contributions employed in this
analysis, is the correlation of particles separated in rapidity by
large interval~\cite{method}.
In particular we present the results for elliptic flow measured in the 
STAR main TPC ($-0.9<\eta<0.9$), obtained via 
correlations with particles in the two Forward TPCs ($2.9<|\eta|<3.9$).
The results presented below are based on an analysis (after
all event quality cuts) of 9.6 M Au+Au 200 GeV, 7 M Au+Au  62 GeV, 30 M
Cu+Cu 200 GeV, and 19 M Cu+Cu  62 GeV   Minimum Bias events
(note the better statistics for 200 GeV Au+Au and Cu+Cu collisions
compared to the previously reported results in~\cite{voloshin:CIPANP}).
Centrality of the collision is determined in accordance with the
so-called Reference Multiplicity - the multiplicity of primary tracks
in $|\eta|<0.5$ region. 

Fig.~1 presents the results for elliptic flow of charged
particles in the pseudorapidity window $|\eta|<0.9$, 
for 200 GeV Au+Au and Cu+Cu collisions. 
The corresponding  62~GeV results can be found in~\cite{voloshin:CIPANP}.
Charged particles are selected from $0.15<p_t<2.0$~GeV region; the low
transverse momentum cut is due to TPC acceptance. 
In black, noted as $v_2\{2\}$, 
are shown the results obtained from two particle azimuthal correlations with 
both particle from the main TPC region. 
In blue, noted as $v_2\{\rm{FTPC}\}$,  are the results obtained 
correlating particles in the main
and Forward TPCs regions. The larger values of  $v_2\{2\}$ compared
to $v_2\{\rm{FTPC}\}$ are attributed to the non-flow contribution.
The relative contribution of non-flow in Cu+Cu collisions is
significantly larger compared to that in Au+Au collisions due to smaller values
of flow itself. 
Analysis (similar to that performed in~\cite{star-flow1})
of the centrality dependence of correlations between
flow vectors obtained in the two Forward TPCs and also with flow vector in
the Main TPC 
yields an estimate of the systematic relative error at maximum flow of the
order of $\leq 3$\%    (AuAu 200 GeV),
$\leq 5$\%     (AuAu 62 GeV),
$\leq 12$\%  (CuCu 200 GeV), and 
$\leq 20$\%  (CuCu 62 GeV).

\begin{figure}
  \includegraphics[width=.55\textwidth]{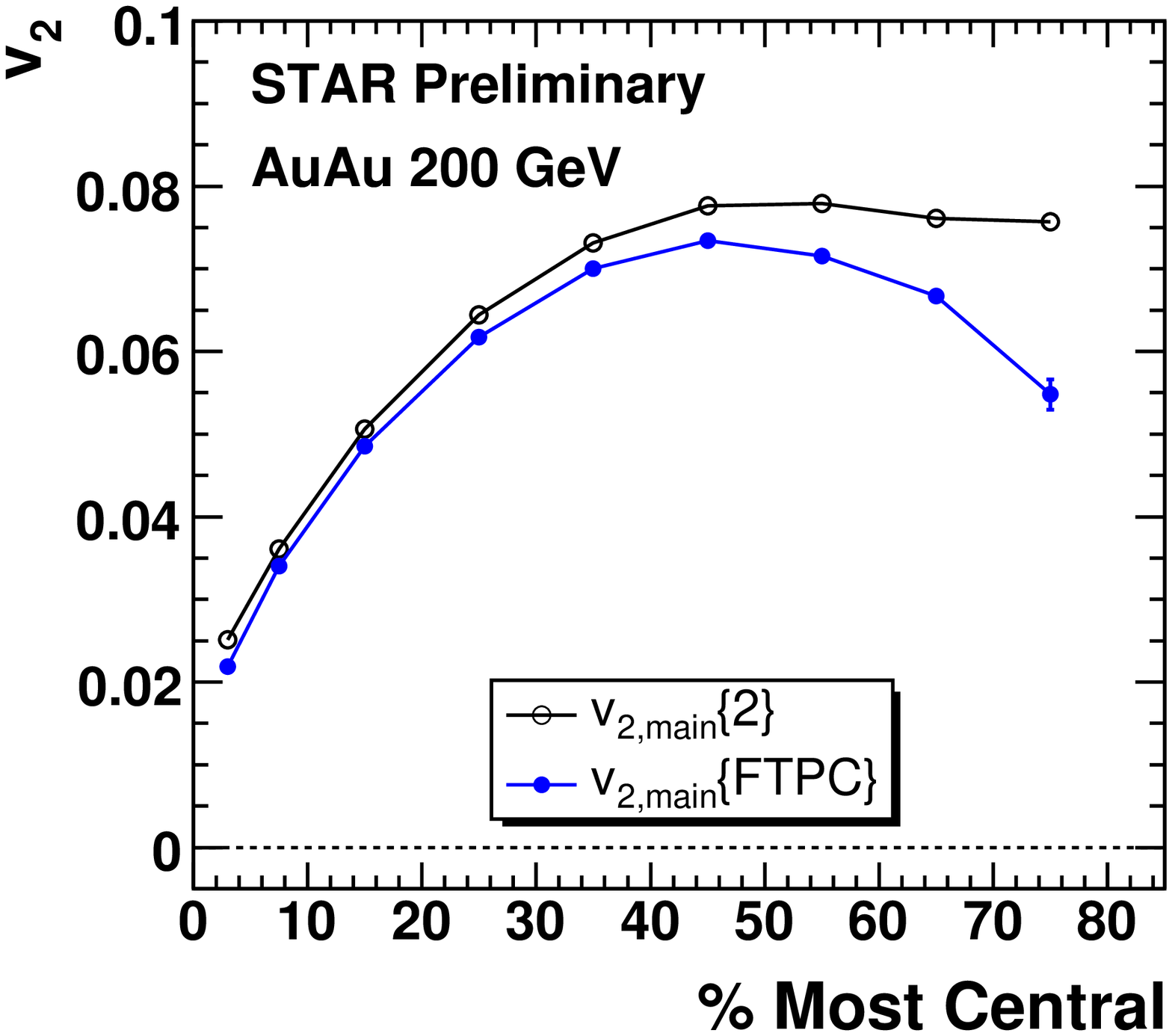} \hspace{-0.05\textwidth}
  \includegraphics[width=.55\textwidth]{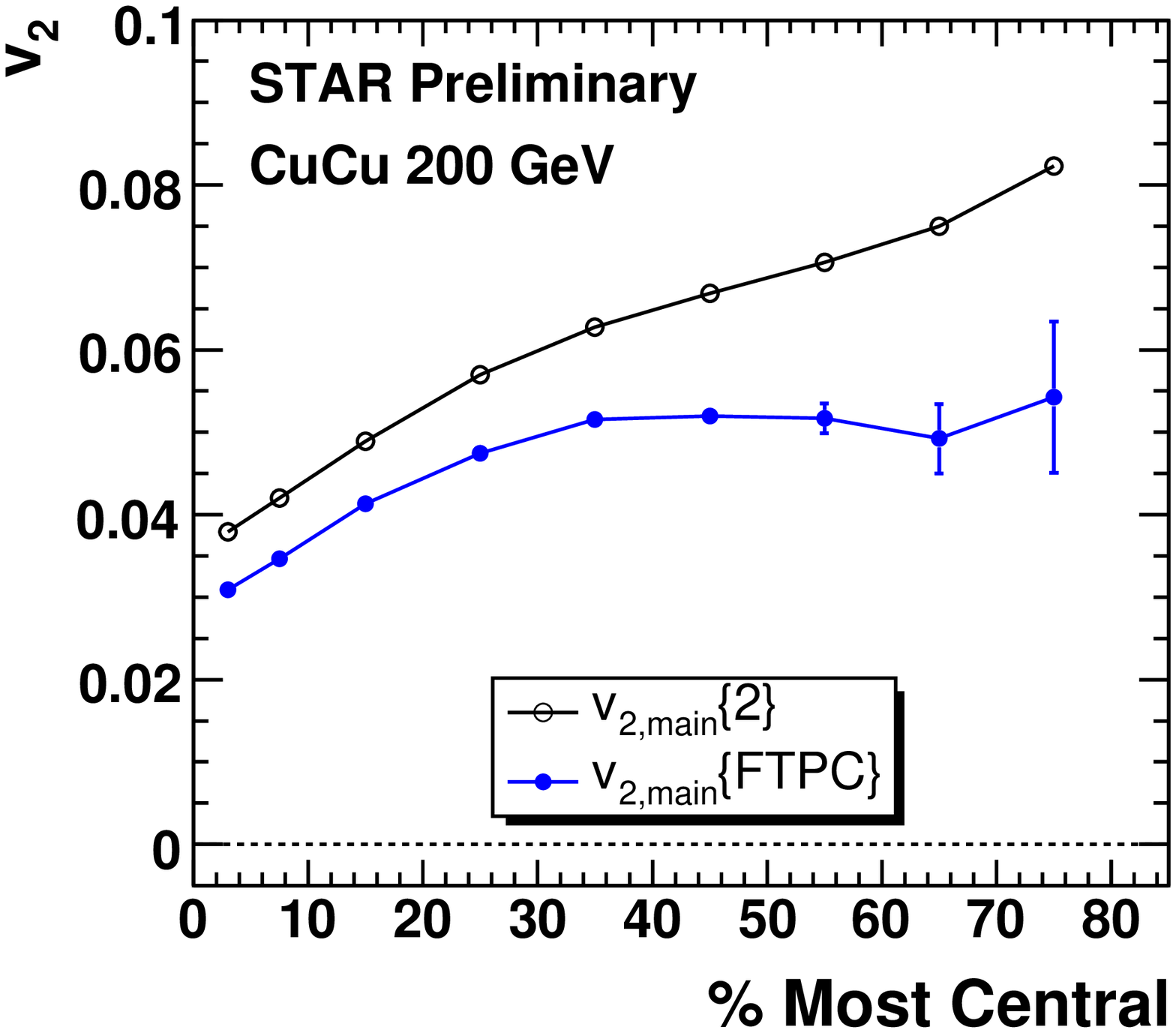}
  \caption{ (color online) Elliptic flow in Au+Au and Cu+Cu  collisions 
    as function of centrality at $\sqrt{s_{NN}}=$200 GeV. 
}
\end{figure}

\begin{figure}
  \includegraphics[width=.55\textwidth]{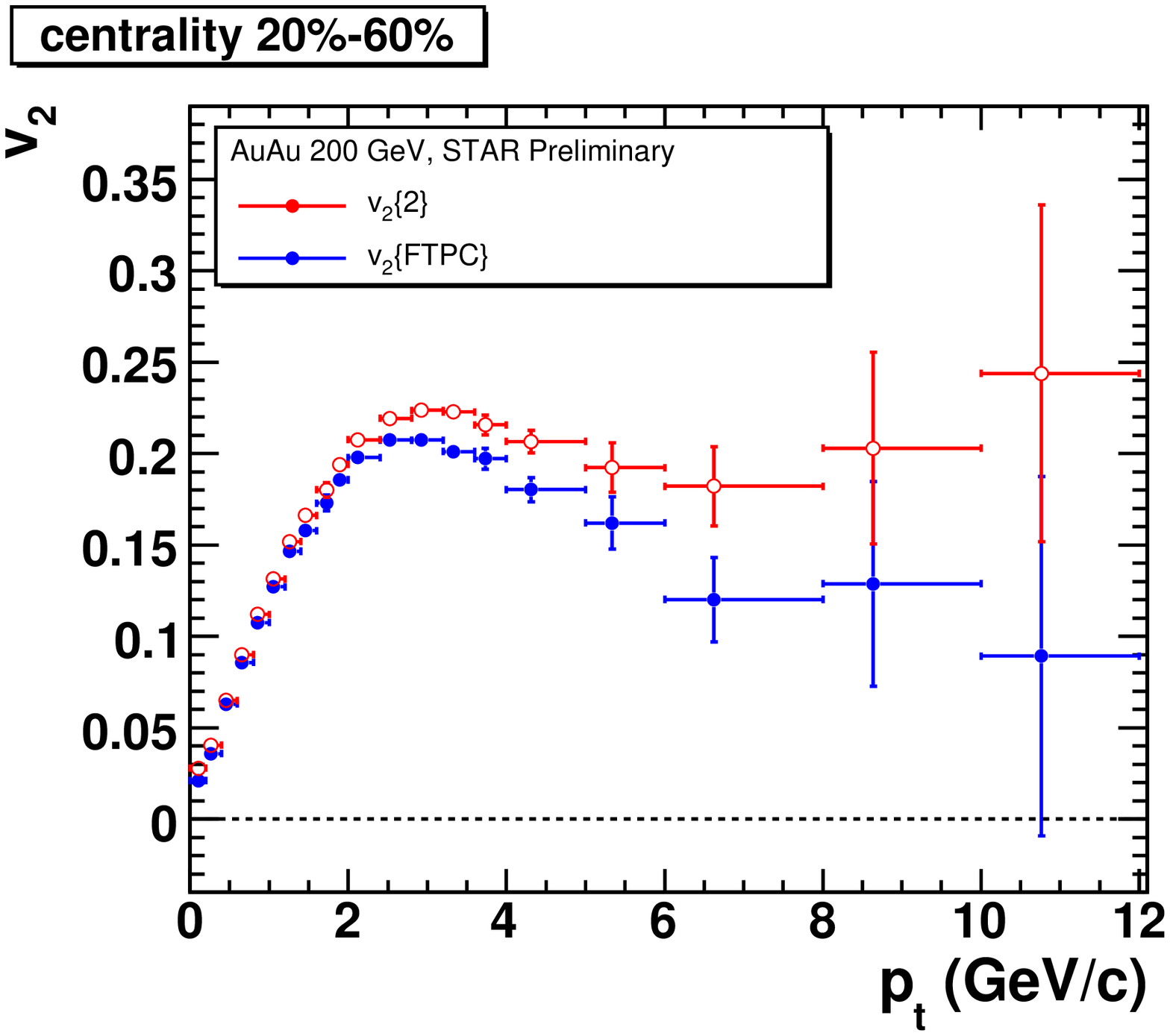}
	\hspace*{-0.8cm}
  \includegraphics[width=.55\textwidth]{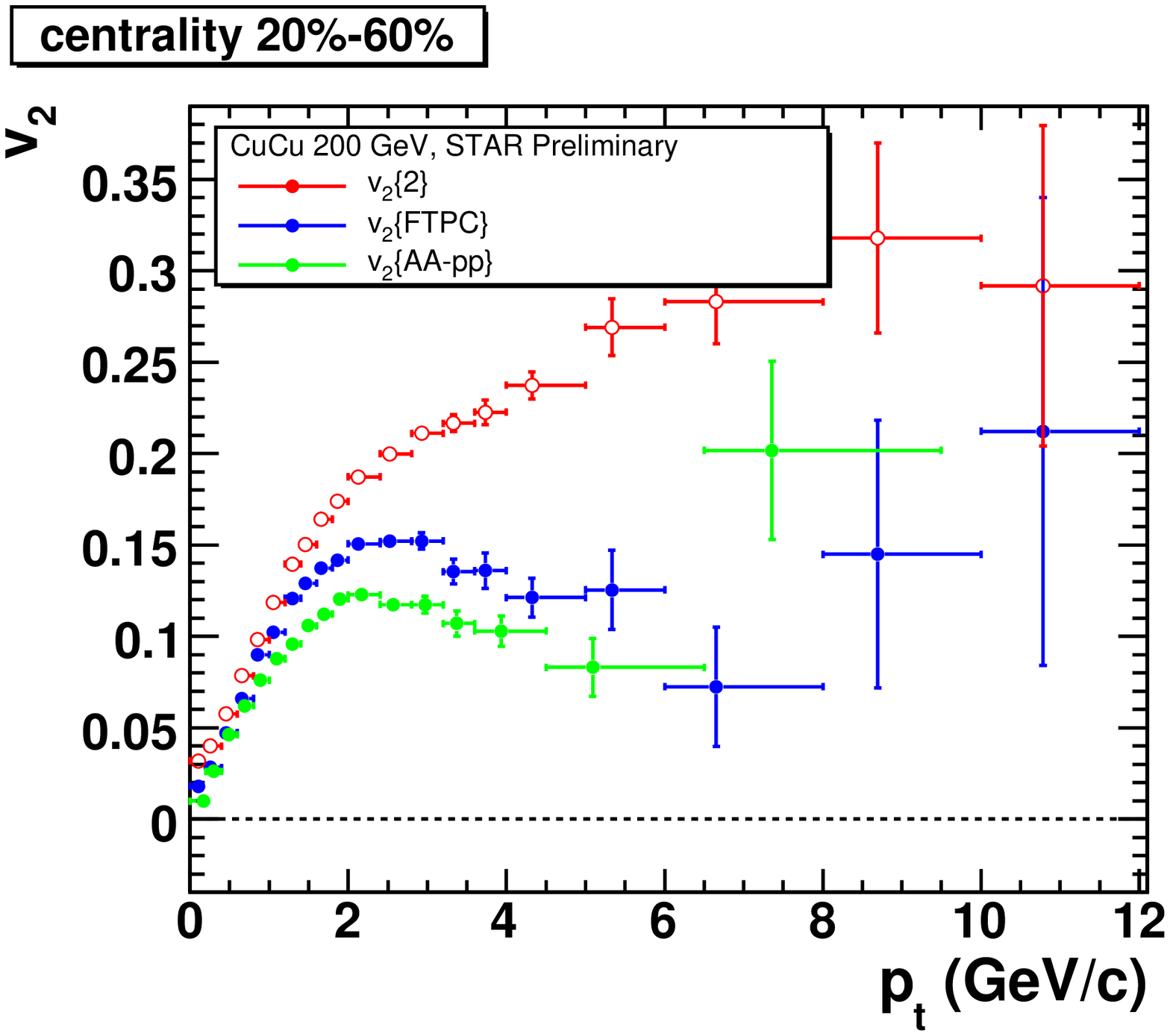}
  \caption{(color online) 
$v_2(p_t)$ in midcentral Au+Au and Cu+Cu collisions at
$\sqrt{s_{NN}}=$200 GeV.}
\label{fv2pt}
\end{figure}

Figure~\ref{fv2pt} presents the results for $p_t$ dependence of elliptic
flow in Au+Au and Cu+Cu collisions at $\sqrt{s_{NN}}=200$~GeV.
Two sets of results are shown: $v_2\{2\}$ -- two
particle correlation results with both particles in the main TPC
region, and $v_2\{FTPC\}$.
Cu+Cu plot additionally shows $v_2\{AA-pp\}$ which is 
$v_2\{2\}$ with subtracted non-flow contribution as measured in
$pp$-collisions. Differential flow results indicate that non-flow
contribution increases with transverse momentum, and at $p_t>6$~GeV
non-flow effects can become significant. 
$v_2(p_t)$ calculated 
at different collision centralities (not presented in the proceedings)
indicate relatively larger non-flow effects at high transverse momentum
in more central collisions.

Adequate treatment of
flow fluctuations is essential for understanding the flow dependence on
centrality of the collision and the size of colliding nuclei. 
Under the assumption that elliptic flow
closely follows initial eccentricity of the system, 
$\eps=\mean{y^2-x^2}/\mean{y^2+x^2}$, 
one can estimate flow fluctuations by calculation
of eccentricity fluctuations~\cite{star-flow-PRC,raimond-mike}, 
e.g. in Glauber Monte-Carlo model. It was
argued~\cite{manly-qm2005,voloshin:LaJolla} that the so-called
``participant'' eccentricity should be used in this calculations.
The corresponding results for the four systems studied in this analysis
have been presented
in~\cite{voloshin:LaJolla}, and are used in the current analysis.
The direct measurement of flow fluctuations presented in this 
conference~\cite{sorensenQM2006} are consistent with the assumption
that flow fluctuations are dominant by fluctuation in the initial
eccentricity. 

\begin{figure}
  \includegraphics[width=.66\textwidth]{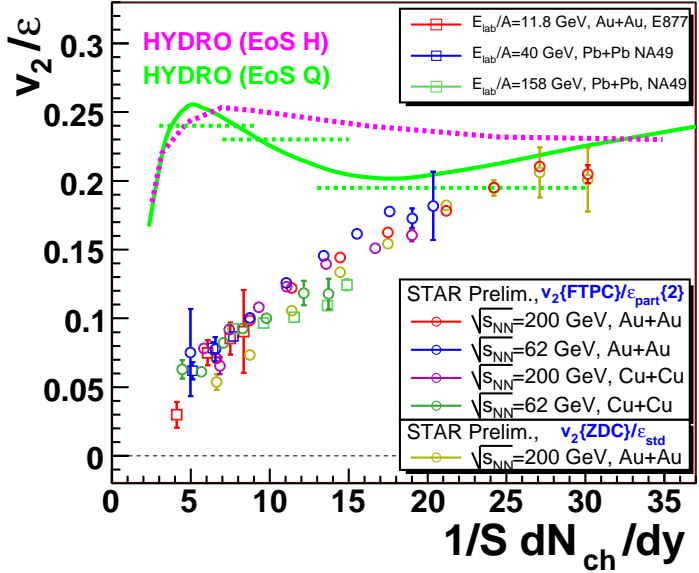}
  \vspace{-5mm}
  \caption{(color online) $v_2/\eps$ scaling plot}
\label{fv2eps}
\end{figure}

Fig.~\ref{fv2eps} shows how new
results on the integrated elliptic flow fit to the $v_2/\eps$ scaling. 
This plot is made under assumption that flow
fluctuations in the main TPC region, $|\eta|<0.5$, are fully correlated with
flow fluctuations in the Forward TPC regions $2.9<|\eta|<3.9$, 
which justifies rescaling with $\sqrt{\mean{\eps_{part}^2}}$. 
(More details on how flow results obtained in different
ways should scale with eccentricity can be found
in~\cite{voloshin:LaJolla}.)
The results obtained with the use of the first order reaction plane
from STAR ZDC-SMD~\cite{gangQM05} are rescaled with
$\mean{\eps_{part}}$, as they are not a subject to flow fluctuations. 
For the references to other data presented in Fig.~3 see~\cite{na49-flow-PRC}. 

Note that for the  $v_2/\eps$ scaling plot the results from Figures 1
and 2 have been
rescaled/extrapolated to a full transverse momentum coverage using
blast wave fits to spectra; rapidity density has been obtained from
pseudorapidity density using scaling factors based on HIJING and RQMD
calculations. 
The presented (ideal) hydrodynamic predictions are based on
calculations~\cite{hydro}. The curves shown are obtained from hydro
results made for fixed impact parameter ($b=7$~fm) and different
particle densities (collision energies). 
Note that hydro results do not scale perfectly in
this plot and in general exhibit somewhat flatter centrality
dependence at each collision energy.

Fig.~\ref{fv2eps} shows that, despite that  62 GeV Au+Au results are somewhat
higher than the rest, in general the
scaling~\cite{voloshin-poskanzer-PLB} holds well.
Also remarkable is the agreement between $v_2\{FTPC\}$ and
$v_2\{ZDC-SMD\}$ rescaled with appropriate (but different) values of 
eccentricity.
Obviously the LHC and/or RHIC U+U data will be very valuable to
understand what happens at even higher particle density.


\end{document}